\documentclass[mathleft
]{an}
\usepackage{graphicx}
\usepackage{times}
\begin{document}

\Yearpublication{2010}%
\Volume{331}%
\Issue{9-10}%
 \DOI{P58}%

\title{MISOLFA solar monitor for the ground PICARD program }

\author{T. Corbard\inst{1}\fnmsep\thanks{Corresponding author:
  \email{Thierry.Corbard@oca.eu}\newline}
\and  A. Irbah\inst{2}
\and  P. Assus\inst{1}
\and C. Dufour\inst{2}
\and M. Fodil\inst{3}
\and F. Morand\inst{1}
\and C. Renaud\inst{1}
\and E. Simon\inst{1}
}
\titlerunning{MISOLFA solar monitor for the ground PICARD program}
\authorrunning{T. Corbard et al.}
\institute{
Universit\'e de Nice Sophia Antipolis, CNRS, Observatoire de la C\^ote d'Azur,
BP 4229 06304 Nice Cedex 4, France
\and 
Universit\'e Versailles St-Quentin, CNRS/INSU, LATMOS-IPSL, Guyancourt, FRANCE
\and 
CRAAG, Observatoire d'Alger, BP 63 Bouzar\'eah 16340 Alger, Alg\'erie
}



\abstract{%
Developed at the Observatoire de la C\^ote d'Azur (OCA) within the framework of the 
PICARD space mission (Thuillier et al., 2006) and with support from the french 
spatial agency (CNES),  MISOLFA (Moniteur d'Images Solaires Franco-Alg\'erien) is
a new generation of daytime turbulence monitor. 
Its objective is to measure both the spatial and temporal turbulence parameters in
order to quantify their effects on the solar diameter measurements that will be made from
ground using the qualification model of the SODISM (SOlar Diameter Imager and 
Surface Mapper) instrument onboard PICARD.
The comparison of simultaneous images from ground and space should allow us,
with the help of the solar monitor, to find the best procedure possible to measure 
solar diameter variations from ground on the long term. 
MISOLFA is now installed at the Calern facility of OCA and PICARD is scheduled
to be launched in 2010. 
We present here the principles of the instrument and the first results obtained
on the characteristics of the turbulence observed at Calern observatory
using this monitor while waiting for the launch of the space mission. 
}

\maketitle

\section{Introduction: why a new Solar monitor ?}
\begin{figure}
\includegraphics[width=\linewidth]{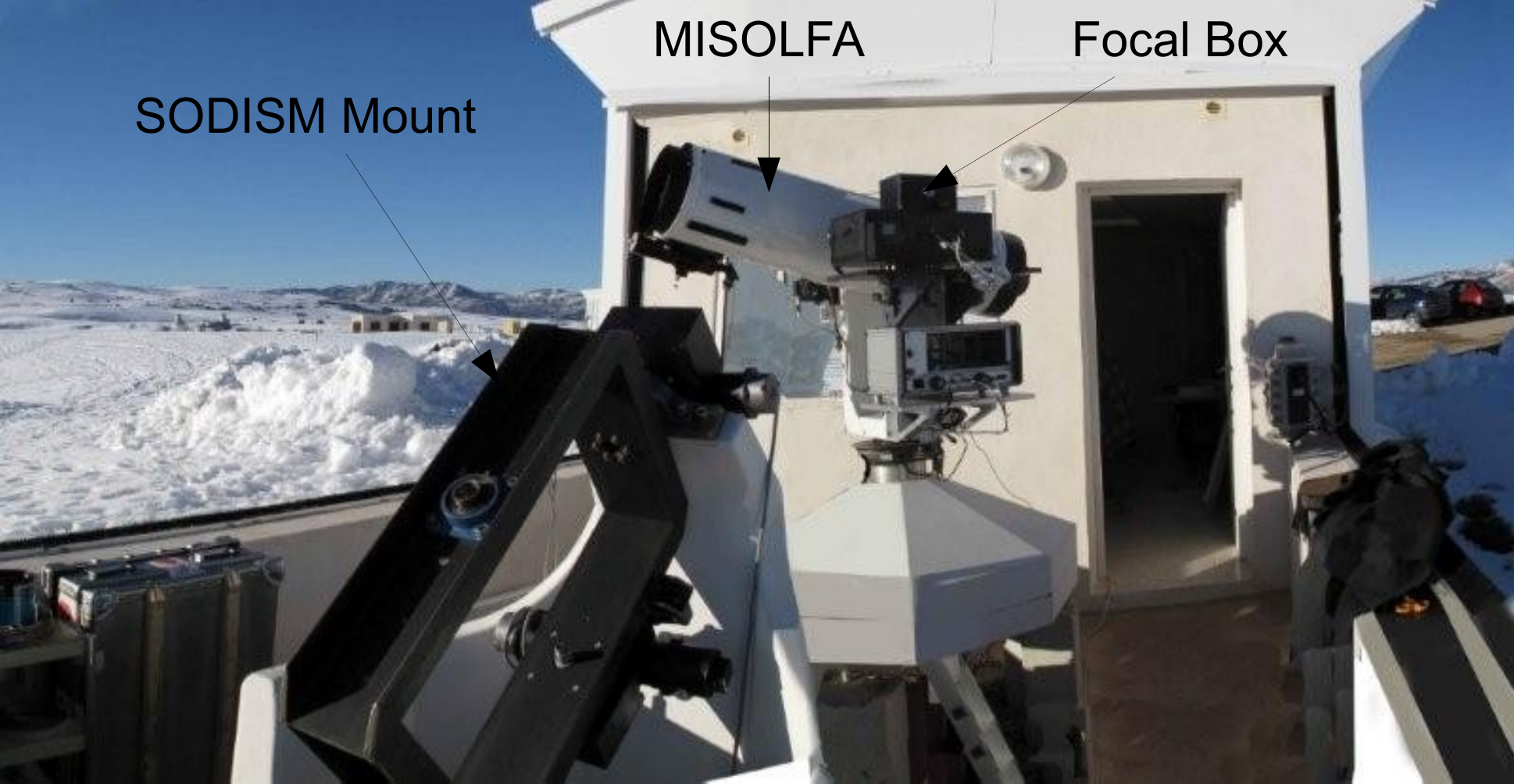}
\caption{Picture of MISOLFA monitor and the mount that will recieve the qualification model 
of the PICARD/SODISM telescope at the Calern facility of OCA.}
\label{fig:photo}
\end{figure}
\begin{figure*}
\includegraphics[width=\linewidth,height=4cm]{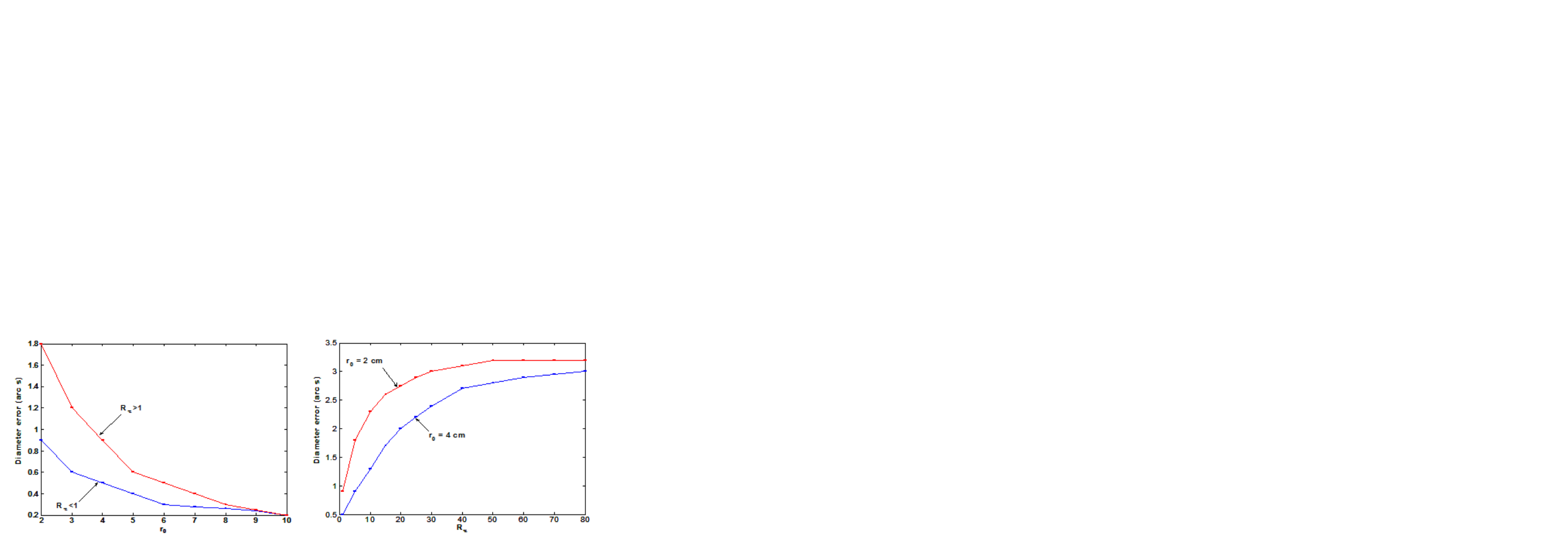}
\caption{Estimated errors on diameter estimates as a function of the Fried's parameter $r_0$
 for two values of $R_{\tau}$ (ratio between the exposure time and the correlation time) 
 (left) and as a function of $R_{\tau}$ for two values of $r_0$ (right) (Lakhal et al., 1999)}
\label{fig:lakhal}
\end{figure*}

Diameter measurement analysis revealed a dependence with the seeing condition 
as represented by Fried's parameter $r_0$ (Irbah et al., 1994). 
Numerical simulations were developed in order to better understand  atmospheric effects
on diameter measurements. The error decreases with the seeing but it is also strongly 
conditioned  by turbulence coherence times (Fig.\ref{fig:lakhal}, Lakhal et al., 1999). The error also 
shows weak dependence on the outer scale $L_0$ for a small aperture telescope. 
Existing solar monitors such as  Solar DIfferential Motion Monitors (S-DIMM)
or arrays of scintillometers (SHABAR) are able to provide useful information 
on the spatial scales of the turbulence and are commonly associated for site testing
(e.g. Beckers 2001). However our goal here is to obtain the accurate turbulence equivalent 
PSF needed to properly interpret the ground based radius measurements. 
For this, estimates of the characteristic temporal scales are also needed and it was realized 
that a single instrument could provide both the spatial and temporal turbulence scales by ana\-lysing Angle of Arrival (AA) fluctuations simultaneously in its image and pupil ways.
     
\section{Atmospheric parameters measured }

An illustration of a multi-layer atmosphere model is given on Fig.~\ref{fig:turb_model}. MISOLFA will allow us to obtain information on all the following parameters using a Von K\'arm\'an turbulence model:

\begin{figure}
\includegraphics[width=\linewidth]{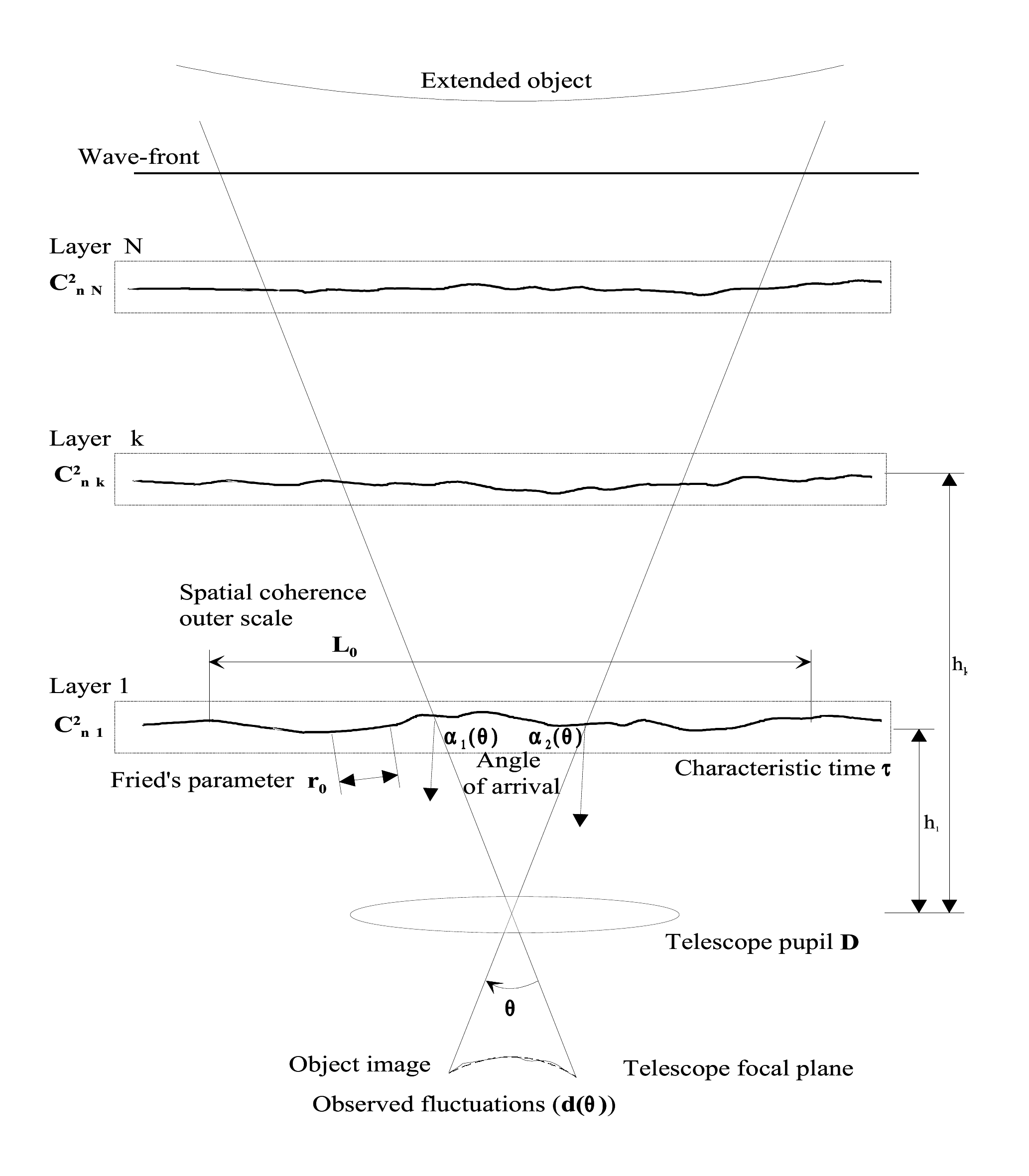}
\caption{Illustration of a multi-layers atmosphere with Von K\'arm\'an turbulence model (Irbah et al., 2003).}
\label{fig:turb_model}
\end{figure}

\begin{itemize}
\item The Atmospheric structure constant of the air refractive index fluctuations $C^2_n(h)$.
\item Fried's parameter $r_0$ which is the diameter of the coherence zone of the degraded wavefront. It corresponds also to the image resolution obtained with the telescope of diameter $r_0$ placed outside the atmosphere. 
\item The spatial coherence outer scale $L_0$ which defines the maximal size of wavefront perturbations remaining coherent. It traduces the low frequency evolution of the wavefront. 
\item The isoplanatic patch $\theta_0$ which is the angle where AA or speckles remain correlated.
\item The correlation time $\tau_0$ which is the time during which the atmosphere may be considered as frizzed for the considered structures (AA, speckles) i.e. the time during which they keep their coherence. 
\end{itemize}

\section{Measurement principles}

The MISOLFA principle is based on the statistical analysis of AA fluctuations,
 which are fluctuations at each point of the normal of the perturbed wavefronts. 
 The AA fluctuations can directly be observed in the image plane
  (case of Shack-Hartmann's sensors used in adaptive optics) but also in 
the pupil plane if the observed source present an intensity distribution with 
a strong discontinuity like the Solar limb. The intensity fluctuations observed in the pupil image are, at first order, proportional to AA fluctuations (Borgnino \& Martin 1977, Borgnino 1978).

\section{MISOLFA Instrument design}

\begin{figure}
\includegraphics[width=\linewidth]{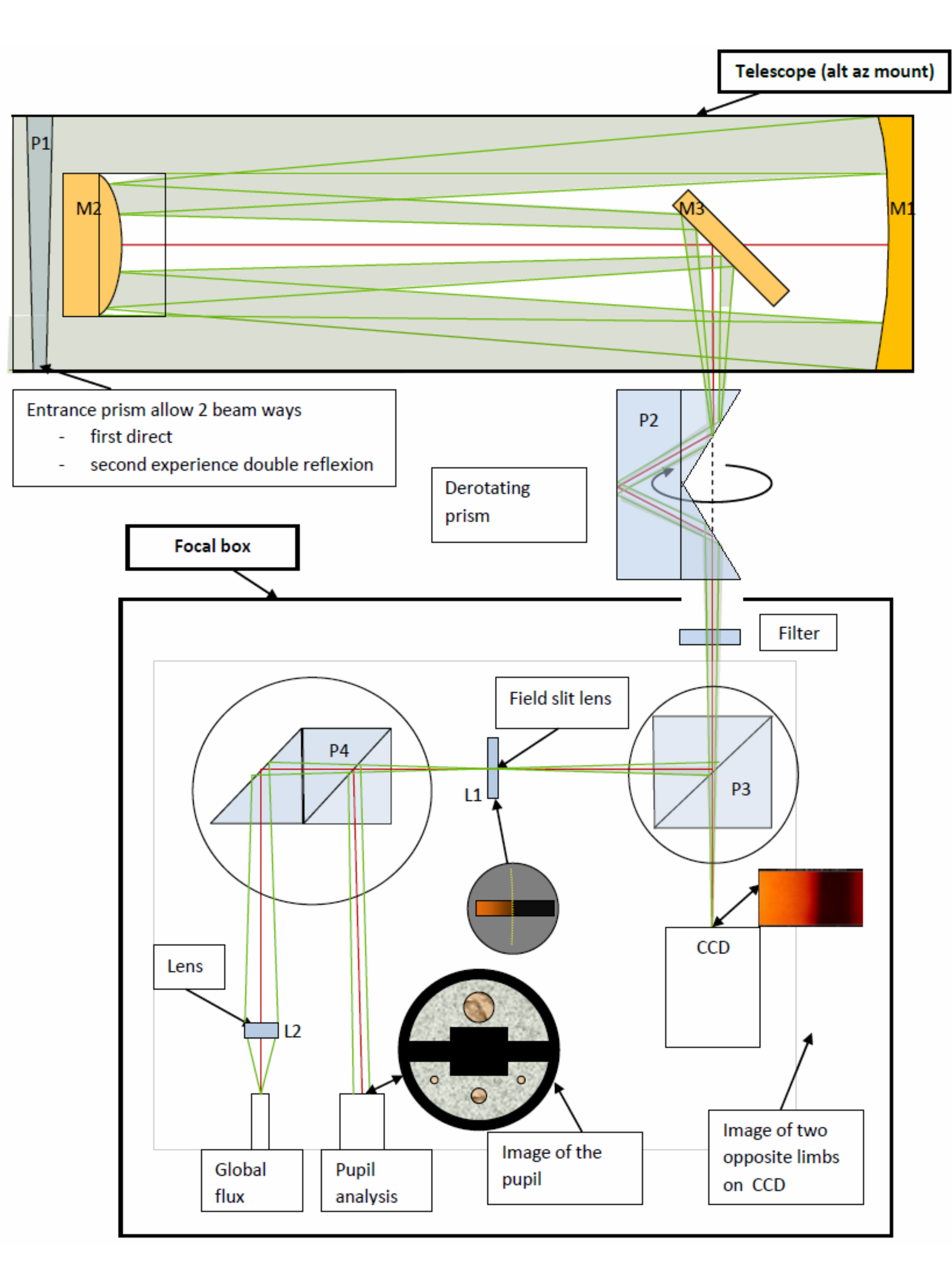}
\caption{MISOLFA optical design. See text for details.}
\label{fig:misolfa_schema}
\end{figure}
\begin{figure}
\begin{center}
\includegraphics[width=0.3\linewidth]{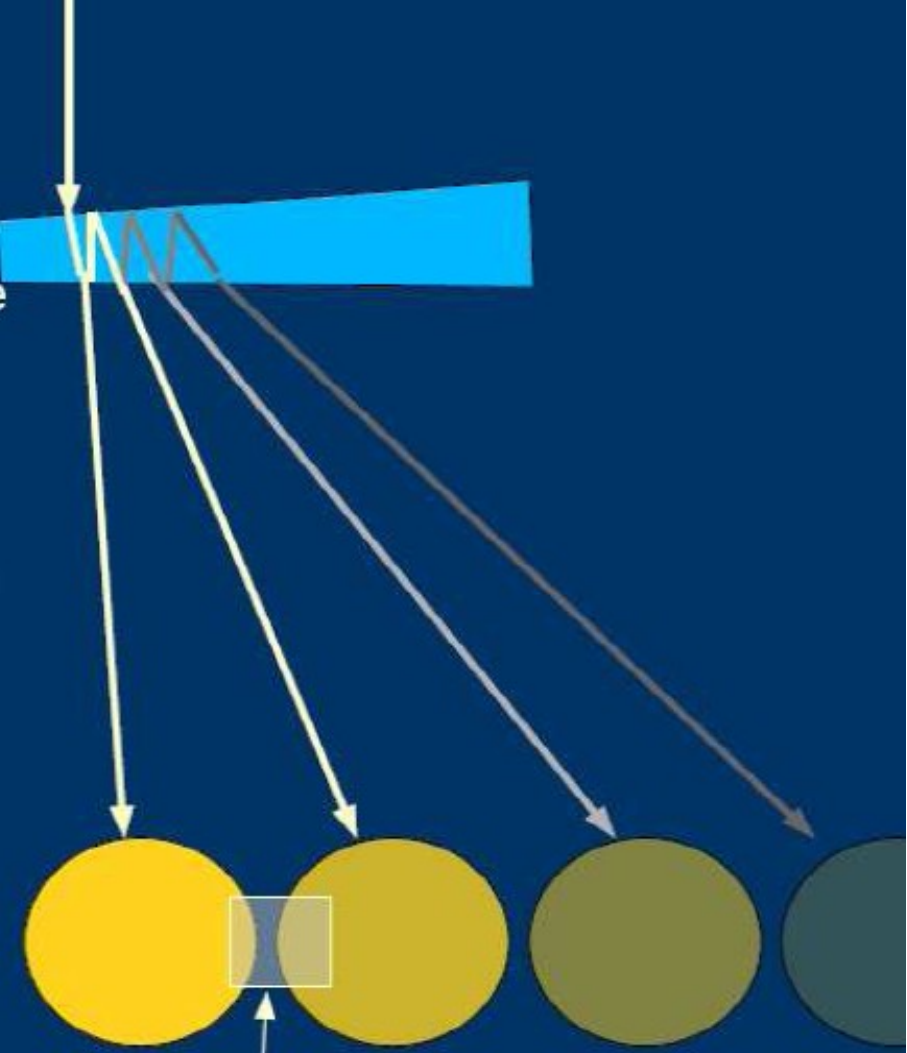}
\end{center}
\caption{Illustration of the formation of several solar images by the entrance prismatic blade. The CCD camera is positioned accross 
the two first solar images.}
\label{fig:lame}
\end{figure}
Figure~\ref{fig:misolfa_schema} illustrates the optical design of MISOLFA.
It is made of a Cassegranian telescope of 25 cm aperture equiped with an alt-azimuth mount to which is associated a derotating system.
A flat glass blade with non parallels faces allows the formation of  separate 
images of the sun (Fig.~\ref{fig:lame}). In the focal box, a first beam (image way) is imaging two opposite parts of solar limb on a 480x640 CCD. A tipycal sequence is $1.5$\,mn long with 32 images per second and a resolution of  0.2 arcseconds per pixel. A second beam (pupil way) is imaging the pupil after passing through a spatial filter (slit) in the image of a part of solar limb. In the pupil image 4 optical fibers (with diameters $2$\,mm, $1$\,mm and 2x0.5\,mm) guide the light to 4 diodes. 
In addition, in the image of the slit a $2$\,mm fiber measure global flux passing through the slit.

\section{Analysis Method}
\begin{figure}
\includegraphics[width=\linewidth,height=4cm]{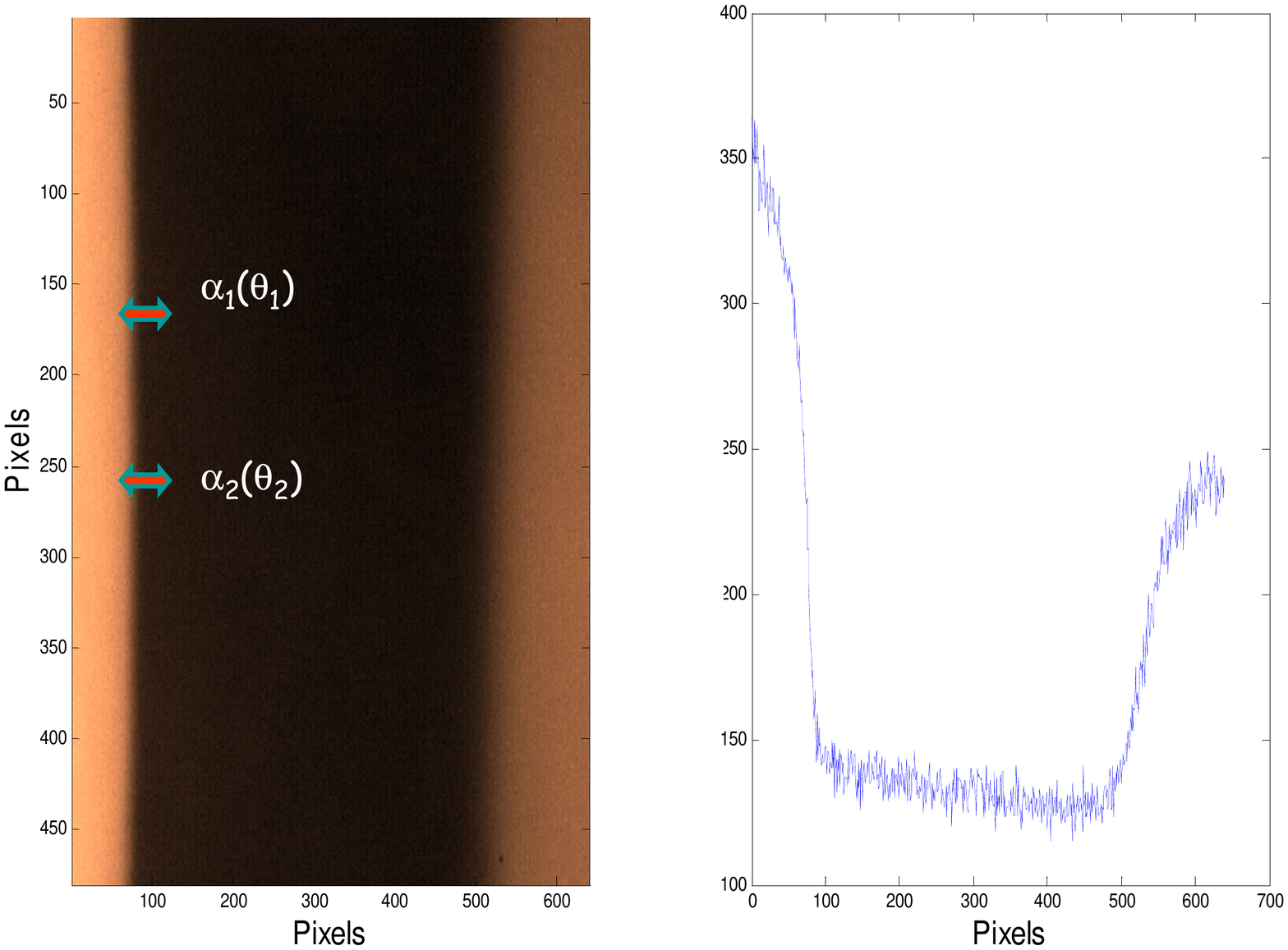}
\caption{Image sample from the MISOLFA monitor (left) and horizontal cut (right). Arrows illustrate
the motions used to compute the transverse covariance between to angular positions (see text).}
\label{fig:cov}
\end{figure}

In the image way we estimate  $r_0$, $L_0$ and $\theta_0$  from direct measurement of  AA fluctuations on the limb images:
\begin{itemize}
\item $r_0$ is directly related to the variance $\sigma_s^2$ of the mean AA fluctuations by:
$$r_0=8.25\-10^5 \ D^{-1/5}\lambda^{6/5}(\sigma_s^2)^{-3/5}$$
where $D$ is the pupil diameter and $\lambda$ the wavelength.
\item The AA transverse covariance is computed  from the observed limb fluctuations  by (Fig.~\ref{fig:cov}): 
$$ C_{\alpha}(\theta)=<\alpha_1(\theta_1)\alpha_2(\theta_2)> $$
$L_0$ and $\theta_0=r_0/h$ are then obtained by adjusting the observed covariance to the theoretical one given, in the case of the one-layer model and Von K\'arm\'an theory, by:
{\setlength{\mathindent}{0pt}
\begin{eqnarray*}
C_\alpha(\theta) &= 0.0716\ \lambda^2\, r_0^{-\frac{5}{3}} \displaystyle\int_0^{\infty} \!\! f^3\left(f+ {{1}\over{L_0^2}} \right)^{-\frac{11}{6}} \ldots\\
 \ldots & \left[J_0(2\pi f \theta h)+J_2(2\pi f \theta h)\right]\left[{\frac{2J_1(\pi D f)}{\pi D f}}\right]^2\mathrm{d}f
\end{eqnarray*}
}
where  $h$ is the altitude of the equivalent impulse layer giving at ground level the same optical effects than 
the whole turbulent terrestrial atmosphere.
\item The turbulence profile $C_n^2(h)$ is then obtained by inverting (Bouzid et al., 2000) the structure function which, in a multi-layer model, is given by:
{\setlength{\mathindent}{0pt}
\begin{eqnarray*}
D_\alpha(\theta) &= &2\left[C_\alpha(0)-C_\alpha(\theta)\right] \\
&=&2.4 \displaystyle\int_0^{\infty} C^2_n(h)K(\theta,h,D,L_0)\mathrm{d}h
\end{eqnarray*}
}
with:
{\setlength{\mathindent}{0pt}
\begin{eqnarray*}
 K(\theta,h,D,L_0)= \displaystyle\int_0^{\infty} \!\! f^3\left(f+ {{1}\over{L_0^2}} \right)^{-\frac{11}{6}} \ldots\\
 \ldots  \left[1-J_0(2\pi f \theta h)-J_2(2\pi f \theta h)\right]\left[{\frac{2J_1(\pi D f)}{\pi D f}}\right]^2\mathrm{d}f
\end{eqnarray*}
}

\end{itemize}
In the pupil plane, we perform a spatio-temporal analysis of AA fluc\-tua\-tions (Berdja, 2007).
 The  photodiode detectors used represent various collecting surfaces of the pupil-plane. They allow fast recordings (5KHz) and thus an accurate intensity signal time sampling. Such measurement time series will be used to study temporal properties of the diurnal turbulence and continuously estimate the correlation time $\tau_0$ of AA-fluctuations from the estimated temporal covariance assuming the Von K\'arm\'an model and Taylor's hypothesis of frozen turbulence.
Such as for other seeing monitors like the well-known S-DIMM or GSM, dual measurements from photodiode pairs on the same collecting surface will allow also estimate the Fried parameter $r_0$. The needed transverse temporal covariance is obtained by doing $\theta h=v\tau$ in the previous formula, $v$ denoting an equivalent speed (weighted average of the different layer speeds) and $\tau$ a temporal shift (Conan et al., 2000).
The goal of using photodiode detectors with different collecting surfaces is to estimate the spatial coherence outer scale $L_0$ from AA-fluctuation statistics. 

\section{Observations and first results}

\begin{figure}
\includegraphics[width=\linewidth]{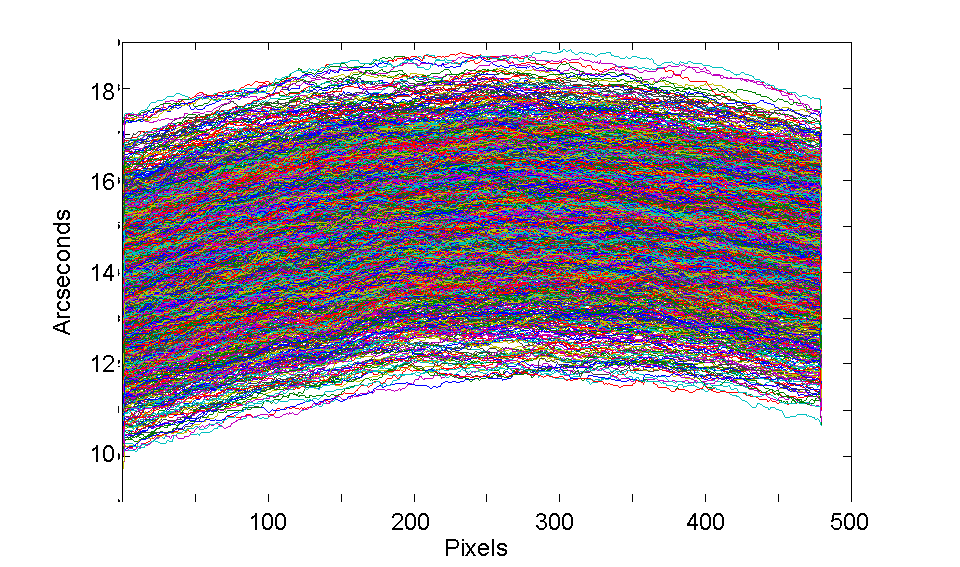}
\caption{The limb profile of each image is extracted using wavelet based denoising and inflexion point identification (Irbah et al. 1999)}
\label{fig:res1}
\end{figure}

\begin{figure}
\includegraphics[width=\linewidth]{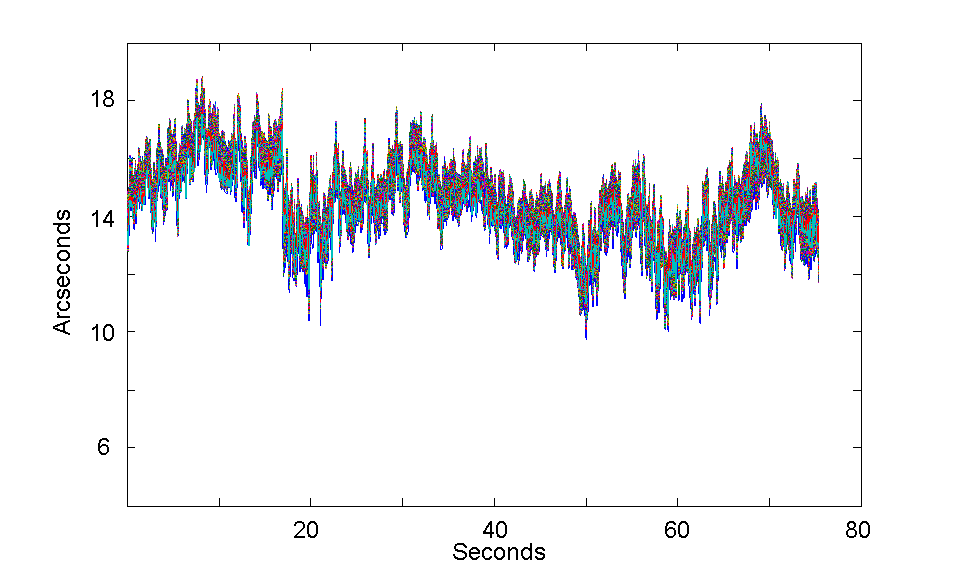}
\caption{Temporal evolution of all the points of the extracted limb profiles. }
\label{fig:res2}
\end{figure}

\begin{figure}
\includegraphics[width=\linewidth,height=4cm]{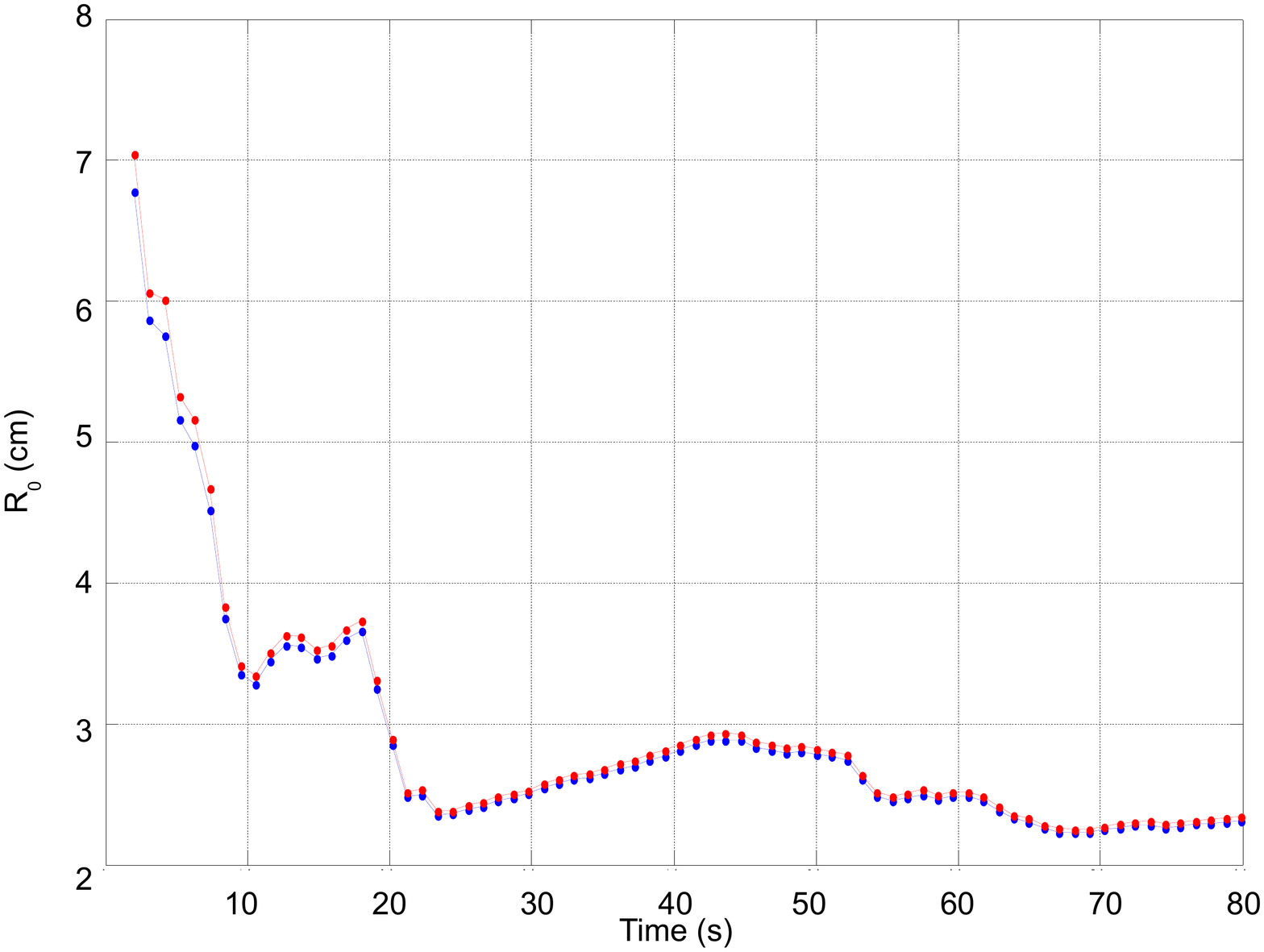}
\caption{Fried's parameter $r_0$ estimated from increasing time intervals. }
\label{fig:res3}
\end{figure}

\begin{figure}
\includegraphics[width=\linewidth,height=4cm]{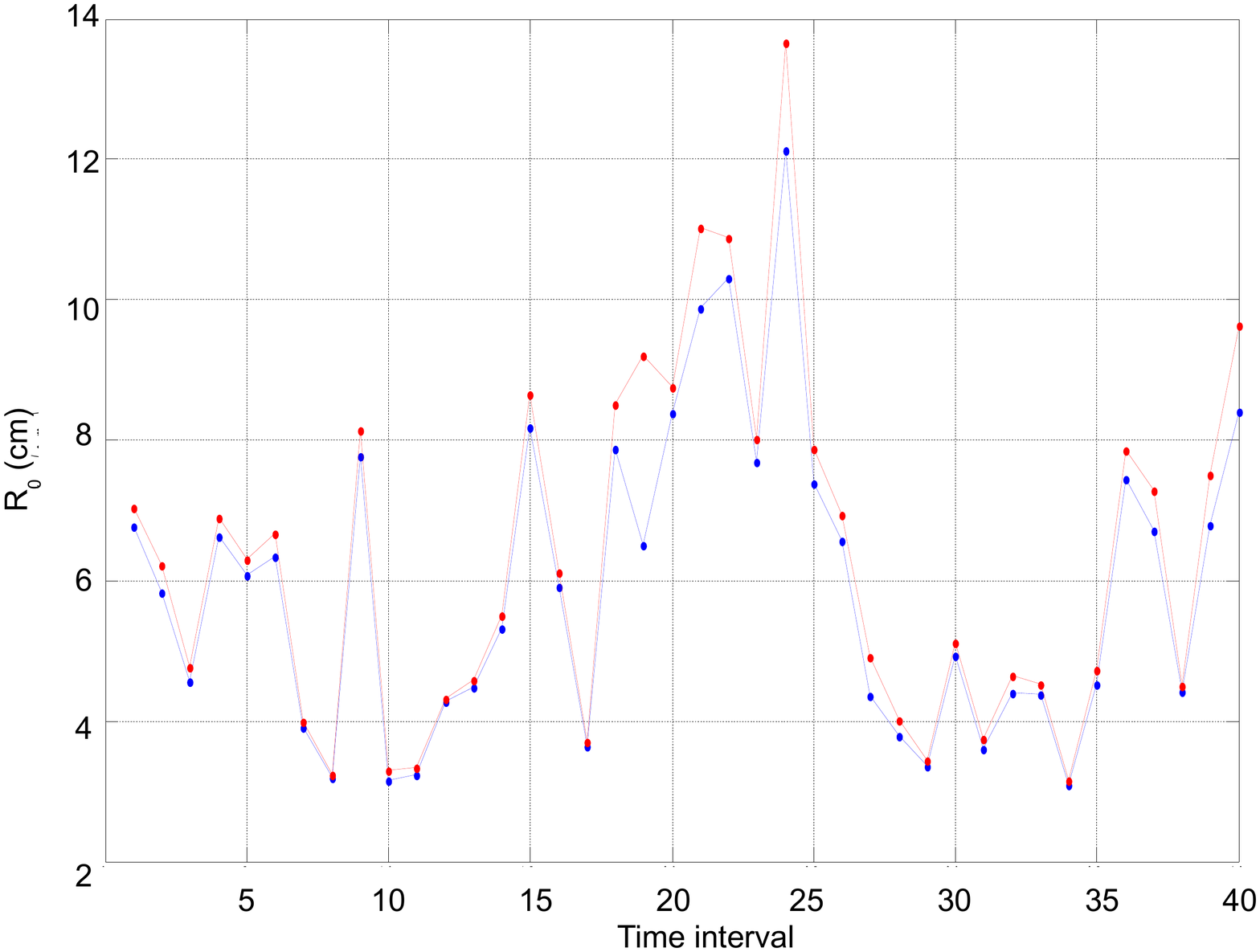}
\caption{Fried's parameter $r_0$ estimated on two seconds intervals. }
\label{fig:res4}
\end{figure}

Sequences of images are recorded at Calern observatory since june 2009. A typical sequence consists in a set of about 2500 images with an angular resolution of 0.2 arcseconds per pixel recorded at a rate of 32 images per second. The exposure time of each image is $1$\,ms. The mean limb position is evaluated for each image in order to correct in real time the guiding errors.  The limb profile of each image is first extracted using wavelet based denoising and inflexion point identification (Irbah et al. 1999) as illustrated Fig.~\ref{fig:res1}. Then, from the temporal evolution of all the points of the extracted limb profiles (Fig.~\ref{fig:res2}),
the variance of the mean AA fluctuations and  the Fried's parameter are estimated. When estimating the Fried's parameter from increasing time intervals, the value decreases while integrating clearly different turbulence regimes (Fig.~\ref{fig:res3}). Figure~\ref{fig:res4} shows Fried's parameter $r_0$ estimated on a continuous set of two seconds intervals which correspond to the exposure time of the full disk images that will be recorded by the SODISM instrument. The full time sequence is therefore cuted in 40 intervals of two seconds. On these time intervals $r_0$ can reach up to $14$\,cm while over the whole $75$\,s the value would be about $2.6$\,cm (Fig.~\ref{fig:res3}).

\section{Conclusion}
MISOLFA is a new type of solar monitor based on the analysis of AA statistic that we record both in the image and pupil plane. It will allow us to get a full characterization of the spatio-temporal parameters of the turbulence needed to identify the appropriate atmospheric model and build the turbulence + instrument equivalent PSF. The first sequences of images obtained where analyzed to extract the Fried's parameter showing that good condition can eventually be reached at Calern observatory on some intervals for the $2$\,s exposure time of SODISM. The continuous record of the $r_0$ values will help us in the interpretation of ground SODISM images while the PICARD satellite is operating. The procedure to extract all the other spatial turbulence parameters have been tested and are now being implemented. The pupil way is still not operating properly and lots of effort are currently made to increase the signal to noise ratio by improving the electronic components.

\acknowledgements
This work has been performed with support of the Observatoire d'Alger - C.R.A.A.G., the French Foreign Affair Ministry, the Observatoire de la C\^ote d'Azur and of the Centre National d'Etudes Spatiales. We thank the European Helioseismology and Asteroseismology Network (HELAS) for supporting our participation to the HELAS-4 conference. HELAS is a major international collaboration funded by the European Commission's Sixth Framework Programme.

\end{document}